\documentclass[12pt,epsfig]{article}
\usepackage{epsfig}
\usepackage{amssymb}
\usepackage{amsmath}
\newcommand{\singlespace}{\renewcommand{\baselinestretch}{1}\large\normalsize}
\newcommand{\doublespace}{\renewcommand{\baselinestretch}{1.6}\large\normalsize}

\newcommand{\beq}{\begin{equation}}
\newcommand{\eeq}{\end{equation}}
\newcommand{\bea}{\begin{eqnarray}}
\newcommand{\eea}{\end{eqnarray}}

\newcommand{\ave}[1]{\langle {#1} \rangle}

\newcommand{\pslash}{p\!\!\!/}

\newcommand{\pb}{\bar\psi}

\newcommand{\eq}[1]{Eq.~(\ref{#1})}
\newcommand{\eqs}[1]{Eqs.~(\ref{#1})}
\newcommand{\ct}{s_{22}}
\newcommand{\cf}{s_{55}}
\newcommand{\cs}{s_{77}}
\newcommand{\xup}{x_{uu}^+}
\newcommand{\xum}{x_{uu}^-}
\newcommand{\xsp}{x_{ss}^+}
\newcommand{\xsm}{x_{ss}^-}
\newcommand{\xusp}{x_{us}^+}
\newcommand{\xusm}{x_{us}^-}
\newcommand{\yu}{y_{uu}}
\newcommand{\ys}{y_{ss}}
\newcommand{\yus}{y_{us}}
\newcommand{\dt}{|\Delta_2|}
\newcommand{\df}{|\Delta_5|}

\def\roughly#1{\mathrel{\raise.3ex\hbox{$#1$\kern-.75em%
\lower1ex\hbox{$\sim$}}}}

\def\gsim{\roughly>}

\def\={\;=\;}
\def\+{\;+\;}

\begin{document}

\begin{flushright}
December 2001
\end{flushright}
\vspace{1.0cm}
\begin{center}
\doublespace
\begin{large}
{\bf } Color-Flavor Unlocking and Phase Diagram with Self-Consistently
Determined Strange Quark Masses\\
\end{large}
\vskip 1.0in
M. Buballa$^{a,b}$ and M. Oertel$^{c}$\\
{\small{\it $^a$ Institut f\"ur Kernphysik, TU Darmstadt,
                 Schlossgartenstr. 9, 64289 Darmstadt, Germany\\
            $^b$ Gesellschaft f\"ur Schwerionenforschung (GSI),
                 Planckstr. 1, 64291 Darmstadt, Germany\\
            $^c$ IPN-Lyon, 43 Bd du 11 Novembre 1918,
                 69622 Villeurbanne C\'edex, France}}\\
\end{center}
\vspace{1cm}

\begin{abstract}
The phase diagram of strongly interacting matter at non-zero temperature and
baryon chemical potential is calculated within a 3-flavor NJL-type quark 
model with realistic quark masses. The model exhibits spontaneous 
chiral symmetry breaking as well as diquark condensation in the 
two-flavor color-superconducting phase and in the color-flavor locked phase. 
We investigate the color-flavor unlocking phase transition, taking into
account self-consistently calculated effective quark masses.     
We find that it is mainly triggered by a first order phase transition with
respect to the strange quark mass. It takes place at much
higher values of the chemical potential than the transition to the
hadronic phase such that we find a relatively large region in the
phase diagram where the two-flavor color-superconductor seems to be
the most favored state.   
\end{abstract}

\newpage

\singlespace

\section{Introduction}
\label{introduction}

The structure of the QCD phase diagram is one of the most exciting topics
in the field of strong interactions (For review see, e.g. 
\cite{Halasz99,Raja99,RaWi00,Alford01}).
For a long time the discussion was 
restricted to two phases: the hadronic phase and the quark-gluon plasma (QGP).
The former contains ``our'' world, where quarks and gluons are confined
to color-neutral hadrons and chiral symmetry is spontaneously broken due to
the presence of a non-vanishing quark condensate $\phi = \ave{\pb\psi}$.
In the QGP quarks and gluons are deconfined and chiral symmetry is
(almost) restored, $\phi\simeq 0$.
 
Although color-superconducting phases were discussed already in the '70s 
\cite{Ba77,Fr78} and '80s \cite{BaLo84}, until quite recently not much 
attention was payed to this possibility.
This changed dramatically after it was discovered that due to 
non-perturbative effects, the gaps which are related to these phases could 
be of the order of $\Delta\sim 100$~MeV \cite{ARW98,RSSV98}, much larger 
than expected from the early perturbative estimates. 
Since in standard weak-coupling BCS theory the critical temperature is given 
by $T_c \simeq 0.57\,\Delta(T=0)$ \cite{FW71}, this also implies
a sizable extension of the color-superconducting phases into the 
temperature direction \cite{PiRi99}.
Hence, color-superconducting phases could be relevant
for neutron stars \cite{We99,BGS01} and -- in very optimistic cases -- 
even for heavy-ion collisions \cite{RiPi00}. 

Many different diquark condensates are allowed by Pauli principle 
\cite{BaLo84}. However, most interactions, like one-gluon exchange  
or instanton mediated interactions favor a condensation in the scalar 
color-antitriplet channel. In general, these condensates read
\beq
    s_{AA'} 
    \= \ave{\psi^T \,C \gamma_5 \,\tau_A \,\lambda_{A'} \,\psi} \;,
\label{saa}
\end{equation}
where both, $\tau_A$ and $\lambda_{A'}$ are the antisymmetric
generators of $SU(N_f)$ and $SU(N_c)$, acting in flavor space and in color 
space, respectively. Throughout this paper, we will restrict ourselves to
the physical number of colors, $N_c=3$. Then the $\lambda_{A'}$ denote the
three antisymmetric Gell-Mann matrices, $\lambda_2$, $\lambda_5$ and 
$\lambda_7$. 

For two flavors ($N_f= 2$), the flavor index in \eq{saa} is restricted to 
$A=2$, describing the pairing of an up quark with a down quark.
The three condensates $s_{2A'}$, $A' = 2, 5, 7$, form a vector in 
color space, which always can be rotated into the $A' = 2$-direction. 
Hence the two-flavor superconducting state (2SC) state can be 
characterized by
\beq
    \ct \neq 0  \quad \text{and}\quad
    s_{AA'} = 0 \quad \text{if}\quad (A,A') \neq (2,2) \,.
\label{2cs}
\end{equation}
Due to the presence of $\ct$, color $SU(3)$ is spontaneously broken down
to $SU(2)$. For massless up and down quarks the 2SC state
is invariant under chiral $SU(2)_L \times SU(2)_R$ transformations. 

For three flavors, the flavor operators $\tau_A$ also denote the
three antisymmetric Gell-Mann matrices, i.e., $A = 2, 5, 7$.  
In this case the two-flavor condensation pattern, \eq{2cs} is still
possible, but now there are several other combinations which cannot be 
transformed into $\ct$ via color or flavor rotations.
If we have three degenerate light flavors, dense matter 
is expected to form a so-called color-flavor locked (CFL) state
\cite{ARW99}, characterized by the situation
\beq
    \ct = \cf = \cs \neq 0  
    \quad \text{and}\quad
    s_{AA'} = 0 \quad \text{if}\quad A\neq A' \,. 
\label{cfl}
\end{equation}
In this state color $SU(3)$ as well as the chiral $SU(3)_L \times SU(3)_R$
and the $U(1)$-symmetry related to baryon-number conservation are broken
down to a common $SU(3)_{color + V}$ subgroup where color and flavor 
rotations are locked.

Both situations discussed so far are idealizations of the real world, where
the strange quark mass $M_s$ is neither infinite, such that strange quarks 
can be completely neglected, nor degenerate with the masses of the up- and 
down quarks.   
However, for sufficiently large quark chemical potentials $\mu \gg M_s$, the 
strange quark mass becomes almost negligible against $\mu$ and a 
phase very similar to \eq{cfl} should form. 
In this case one expects that $\cf$ and $\cs$ are somewhat smaller than 
$\ct$ but do not vanish. Moreover, if we assume an unbroken isospin 
symmetry, $M_u = M_d$, which we will always do in this article, we still
should have $\cf = \cs$. This corresponds to an $SU(2)_{color + V}$ subgroup 
of the original symmetry, where isospin rotations are locked to certain 
$SU(2)$ rotations in color space. Therefore this phase is usually called 
CFL-phase as well.  

At low chemical potentials, the strange quark mass cannot
be neglected against $\mu$ and a phase with $\cf = \cs \neq 0$ is no longer
favored. On the other hand, below a certain value of $\mu$, one 
should finally reach the hadronic phase. 
Hence, if we start in the CFL-phase and keep lowering the chemical potential,
there are (at least) two possible scenarios: 
(1) Below a critical value $\mu_c^{(2)}$, $\cf$ and $\cs$ become zero, 
but $\ct$ remains non-zero, i.e., we have basically
a transition into the 2SC phase (with or without uncondensed strange 
quarks). Upon further decreasing $\mu$ we finally reach the hadronic phase
at some critical chemical potential $\mu_c^{(1)}$. 
(2) The CFL-phase is directly connected to the hadronic phase
without a 2SC-phase in between. This scenario has the interesting feature 
of the so-called quark-hadron continuity, which is related to the fact 
that the symmetries in both regimes are the same and 
the low-lying spectra in both phases can be mapped onto each other   
after the appropriate redefinition of the conserved $U(1)$ charge in the
CFL-phase \cite{ScWi99}. 

It is obvious that the answer to the question which of the two scenarios
is realized in nature depends on the strange quark mass. 
This has first been analyzed by Alford, Berges and Rajagopal \cite{ABR99} 
who have studied the color-flavor unlocking phase transition in a
model calculation with different values of $M_S$. 
Assuming that the region below $\mu \simeq$~400~MeV belongs to the hadronic
phase, these authors came to the conclusion that a 2SC-phase exists if
$M_s \gsim$~250~MeV. Here $M_s$ is an effective ``constituent''mass of
the strange quark, which could be considerably larger than the current
quark mass $m_s \sim$~100 to 150~MeV in the Lagrangian. In general, it is
$T$- and $\mu$-dependent. Moreover, it could depend on the presence of 
quark-antiquark and diquark condensates. In particular, it can be
discontinous along a first-order phase transition line.
This means, not only the phase structure depends on the effective quark 
mass, but also the quark mass depends on the phase. 
This interdependence has not been taken into account in Ref.~\cite{ABR99}, 
where $M_s$ has been kept at fixed values. The authors of
Ref. \cite{rapp00}, who have studied the effect of a nonzero strange
quarks mass within the framework
of an instanton mediated interaction, also neglected these interdependencies. 

The $T$- and $\mu$-dependence of effective quark masses in non-color 
superconducting phases has been studied extensively within NJL-type 
models \cite{vogl,klevansky,hatsuda}, where the masses are closely related 
to the ($T$- and $\mu$-dependent) quark-antiquark condensates
\beq
    \phi_u \= \ave{\bar u u} \= \ave{\bar d d}
    \quad\text{and}\quad
    \phi_s \= \ave{\bar s s}\;.
\end{equation}
Since color-superconductivity is also very often studied within NJL-type 
models, it is the obvious next step to describe both, diquark condensates and
quark-antiquark condensates, simultaneously within the same model. For two 
flavors this has been done by several authors with different degrees of
sophistication \cite{BeRa99,LaRh99,BHO01}. 
For three flavors, a first attempt in this direction was made in 
Ref.~\cite{NGA01}. However, as we will discuss in the next section, the 
thermodynamic potential presented in that article does not have the
correct behavior in the limiting case of a CFL-phase with exact 
$SU(3)$-symmetry and does also not reduce correctly to the 
2SC thermodynamic potential, when $\cf$ and $\cs$ are switched off.\footnote{
In the meantime a substancially revised version of Ref.~\cite{NGA01} appeared,
see footnote on page \pageref{foot2}.}
In the present article we therefore revisit the problem.
In Sect.~\ref{formalism} we present our model and derive the 
thermodynamic potential in mean-field approximation.
Our result has the correct behavior for the 2SC case and for the 
$SU(3)$-symmetric CFL phase. In general, the dispersion laws of the 
corresponding quasiparticle states are more complicated than the result 
presented in Ref.~\cite{NGA01}. In Sect.~\ref{results} we present numerical
results for zero and non-zero temperatures. Finally, conclusions can
be found in Sect.~\ref{conclusions}.

\section{Formalism}
\label{formalism}

We consider the effective Lagrangian
\beq
    {\cal L}_{eff} \= \pb (i \partial\hspace{-2.3mm}/ - \hat{m}) \psi
                      \+ {\cal L}_{q\bar q} \+ {\cal L}_{qq} \,,
\label{Lagrange}
\end{equation}
where $\psi$ denotes a quark field with three flavors and three colors. 
The mass matrix $\hat m$ has the form 
$\hat m = diag(m_u, m_d, m_s)$ in flavor space.
Throughout this paper we will assume isospin symmetry, $m_u = m_d$.

Since we wish to study the interplay between the color-superconducting
diquark condensates $s_{AA'}$ and the quark-antiquark condensates
$\phi_u$ and $\phi_s$, we need an interaction which allows for condensation
in these channels. To this end we consider an NJL-type interaction 
with a quark-antiquark part
\beq
    {\cal L}_{q\bar q} \= G\sum_{a=0}^8 \Big[(\pb \tau_a\psi)^2
    \+ (\pb i\gamma_5 \tau_a \psi)^2\Big]
\label{Lqbarq}
\end{equation} 
and a quark-quark part
\beq
    {\cal L}_{qq} \=
    H\sum_{A = 2,5,7} \sum_{A' = 2,5,7}
    (\pb \,i\gamma_5 \tau_A \lambda_{A'} \,C\pb^T)
    (\psi^T C \,i\gamma_5 \tau_A \lambda_{A'} \, \psi) 
    \,.
\label{Lqq}
\end{equation}
These effective interactions might arise from some underlying more 
microscopic theory and are understood to be used at mean-field level 
in Hartree approximation.

In Eq.~(\ref{Lqbarq}), $\tau_a, a = 1, ..., 8$, denote Gell-Mann matrices 
acting in flavor space, while 
$\tau_0 = \sqrt{\frac{2}{3}}\,1\hspace{-1.5mm}1_f$ is proportional to the 
unit matrix. Hence ${\cal L}_{q\bar q}$ is a $U(3)_L \times U(3)_R$-symmetric 
4-point interaction. In NJL-model studies of the meson spectrum 
\cite{TTKK90,KLVW90} or of properties of quark matter at finite densities or
temperatures \cite{LKW92, Rehberg, BuOe99}, usually a 't Hooft-type 6-point 
interaction is added which breaks the the $U_A(1)$ symmetry.
It is straight forward to take it into account such a term, but for simplicity
we neglect it here. As we will discuss below, this might have 
qualitative consequences for the results.  

In order to calculate the mean-field thermodynamic potential at temperature
$T$ and quark chemical potential $\mu$, we linearize the effective Lagrangian
in the presence of the expectation values $\phi_u=\phi_d$, $\phi_s$, $\ct$, 
and $\cf=\cs$. In this context it is convenient to introduce the effective 
quark masses
\beq
    M_u \= m_u \,-\, 4G\phi_u 
    \qquad\text{and}\qquad
    M_s \= m_s \,-\, 4G\phi_s 
\label{Mus}
\end{equation}
and the diquark gaps
\beq
    \Delta_2 \= -2H \ct\qquad\text{and}\qquad
    \Delta_5 \= -2H \cf~.
\label{Delta12}
\end{equation}
Then, after formally doubling the degrees of freedom,
\beq
     q(x) \;:=\; \frac{1}{\sqrt 2}\,\left(\begin{array}{c} \!\!\psi(x)\!\!\\
     \!\!C\pb^T(x)\!\! \end{array}\right)~,
\label{q}
\end{equation}
the thermodynamic potential per volume can then be written as
\begin{alignat}{1}
    \Omega(T,\mu) \= &-T \sum_n \int \frac{d^3p}{(2\pi)^3} \;
    \frac{1}{2}\,{\rm Tr}\;\ln\, \Big(\frac{1}{T}\,S^{-1}(i\omega_n, \vec p)
    \Big)
    \nonumber\\
    &+\, 4G\,\phi_u^2 \+ 2G\,\phi_s^2 
    \+ H\,(|\ct|^2 \+ 2|\cf|^2)\,. 
\label{Omega}
\end{alignat}
The inverse propagator of the $q$-fields at 4-momentum $p$ is given by
\beq
    S^{-1}(p) \= \left(\begin{array}{cc}
    \pslash - \hat{M} + \mu\gamma^0 &
    \Delta_2 \gamma_5\tau_2\lambda_2 + 
    \Delta_5 \gamma_5 (\tau_5\lambda_5 + \tau_7\lambda_7) \\
    -\Delta_2^* \gamma_5\tau_2\lambda_2
    -\Delta_5^* \gamma_5(\tau_5\lambda_5+\tau_7\lambda_7)  &
    \pslash - \hat{M} - \mu\gamma^0
    \end{array}\right)\,,
\label{Sinv}
\end{equation}
with $\hat{M} = diag(M_u,M_u,M_s)$. In \eq{Omega}, $S^{-1}$ has to be 
evaluated at $p=(i\omega_n, \vec p)$ where $\omega_n = (2n-1)\pi T$ are 
fermionic Matsubara frequencies.

Taking into account the Dirac structure, color, flavor, and the charge
conjugate components, the $q$-fields are 72 dimensional objects, and the
trace in \eq{Omega} has to be evaluated in this 72 dimensional space. 
A tedious but straight-forward calculation yields:
\begin{alignat}{1}
\frac{1}{2}\,{\rm Tr}\;\ln\, \Big(\frac{1}{T}\,S^{-1}(&i\omega_n,\vec p)\Big)
\nonumber\\
\= 3\ln\Big(\frac{1}{T^4}(&\xup\xum +2\dt^2\yu + \dt^4) \Big)
\nonumber\\
\+ 2\ln\Big(\frac{1}{T^4}(&\xup\xsm +2\df^2\yus + \df^4) \Big)
\+ 2\ln\Big(\frac{1}{T^4}(\xsp\xum +2\df^2\yus + \df^4) \Big)
\nonumber\\
\+\phantom{2} \ln\Big(\frac{1}{T^8}(&\xup\xum\xsp\xsm + 2\dt^2 \xsp\xsm\yu
       + 4\df^2(\xup\xsm+\xsp\xum)\yus 
\nonumber\\ 
      &+ \dt^4\xsp\xsm + 4\dt^2\df^2(\xsp\xusm + \xusp\xsm) 
\nonumber\\ 
      &+ 4\df^4(\xup\xsm+\xsp\xum + 4\yus^2) \phantom{\Big)}
\nonumber\\ 
      &+ 8\dt^2\df^4\ys + 32\df^6\yus \+ 16\df^8)\; \Big) \;,
\label{trlog}
\end{alignat}
where we introduced the abbreviations
\beq
    x_{ff'}^\pm \= (\omega_n\pm i\mu)^2 + \vec p^{\,2} + M_f M_{f'}
    \quad\text{and}\quad
    y_{ff'} = \omega_n^2 + \mu^2 + \vec p^{\,2} + M_f M_{f'}\;.
\end{equation}
With these definitions one finds for the argument of the first
logarithm on the r.h.s. of \eq{trlog}
\beq
    \xup\xum +2\dt^2\yu + \dt^4 \= 
    (\omega_n^2 + {E_u^-}^2)(\omega_n^2 + {E_u^+}^2)
\end{equation}
with
\beq
    E_u^\pm \= \sqrt{(\sqrt{\vec p^{\,2} \+ M_u^2} \pm \mu)^2 + \dt^2}\;.
\end{equation}
These are exactly the dispersion laws of the paired quarks in a 
two-flavor color superconductor and the corresponding Matsubara sums are
readily turned out using the standard relation
\beq
     T \sum_n \ln\Big(\frac{1}{T^2}(\omega_n^2 + \lambda_k^2)\Big) \=
    |\lambda_k| \+ 2  T \ln(1+e^{-|\lambda_k|/T})\,.
\label{Matsu}
\end{equation}
The other terms in \eq{trlog} are in general more complicated.
There are, however, two simplifying limits.
The first one corresponds to a two-flavor color superconductor, together with
unpaired strange quarks. In this case $\Delta_5$ vanishes and 
\eq{trlog} becomes
\begin{alignat}{1}
\frac{1}{2}\,{\rm Tr}\,&\ln\, \Big(\frac{1}{T}\,S^{-1}(i\omega_n,\vec p)
\Big)\Big|_{\Delta_5 = 0} 
\nonumber\\
=\quad 
&4\Big[\ln(\frac{\omega_n^2 + {E_u^-}^2}{T^2}) \,+\,
       \ln(\frac{\omega_n^2 + {E_u^+}^2}{T^2})\Big]
\+ 2\Big[\ln(\frac{\omega_n^2 + {\varepsilon_u^-}^2}{T^2}) \,+\,
(\frac{\omega_n^2 + {\varepsilon_u^+}^2}{T^2}) \Big]
\nonumber \\
\+ &3\Big[\ln(\frac{\omega_n^2 + {\varepsilon_s^-}^2}{T^2}) \,+\,
(\frac{\omega_n^2 + {\varepsilon_s^+}^2}{T^2}) \Big]~,
\label{trlog2cs}
\end{alignat}
with $\varepsilon_f^\pm = \sqrt{\vec{p}^{\,2} + M_f^2} \pm \mu$.
Here we recover the fact, that only four of the six light quarks (two colors)
participate in the 2SC condensate, while the two remaining ones 
(and of course all strange quarks) fulfill the dispersion laws of free 
particles with effective masses $M_f$.

We can also reproduce the structure of the dispersion laws of the idealized
3-flavor symmetric CFL-state. 
To this end we evaluate \eq{trlog} for 
$M_u=M_s$ and $\Delta_2 = \Delta_5$. One finds
\begin{alignat}{1}
\frac{1}{2}\,{\rm Tr}\,&\ln\, \Big(\frac{1}{T}\,S^{-1}(i\omega_n,\vec p)
\Big)\Big|_{M_u=M_s,\,\Delta_2=\Delta_5} 
\nonumber\\ 
&\= 8 \Big[\ln(\frac{\omega_n^2 + {E_{oct}^-}^2}{T^2}) \,+\,
           \ln(\frac{\omega_n^2 + {E_{oct}^+}^2}{T^2}) \Big]
\+  \Big[\ln(\frac{\omega_n^2 + {E_{sing}^-}^2}{T^2}) \,+\,
         \ln(\frac{\omega_n^2 + {E_{sing}^+}^2}{T^2}) \Big]
\label{trlog3cfl}
\end{alignat}
with $E_{oct}^\pm = E_u^\pm$ and
$E_{sing}^\pm = \sqrt{(\sqrt{\vec{p}^{\,2} \+ M_u^2} \pm \mu)^2 + 
|2\Delta_2|^2}$. 
We see that in this limit all quarks participate in a diquark condensate,
forming an octet with diquark gap $\Delta_2$ ($=\Delta_5$) and a singlet 
with diquark gap $2\Delta_2$. 
In fact, it has been shown that for three degenerate flavors, the diquark 
gaps form a singlet with $\Delta_{sing} = 2\Delta_{\bar 3} + 4\Delta_{6}$ and
an octet with $\Delta_{oct} = \Delta_{\bar 3} - \Delta_{6}$, 
where ${\bar 3}$ and $6$ refer to pairing in the color-antitriplet and 
color-sextet channel, respectively \cite{ARW99,Sch00,Sho99}. Since we have 
neglected the small color-sextet contribution in our ansatz, we should find
$\Delta_{sing} = 2\Delta_{oct}$, in agreement with the above results.

Both limiting cases, \eqs{trlog2cs} and (\ref{trlog3cfl}), are not
correctly reproduced, if one starts from the thermodynamic potential
given in Ref.~\cite{NGA01}. For instance, there are always quarks which 
fulfill free dispersion laws (no diquark gaps), which should not
be the case in the CFL-phase, see above.\footnote{
After submission of our final manuscript, Ref.~\cite{NGA01} was 
replaced by a substancially revised version ({\it F. Gastineau et al., 
hep-ph/0101289 v3}). The quark dispersion laws presented there
are now consistent with the limits \eqs{trlog2cs} and (\ref{trlog3cfl}). 
The authors make the simplifying assumption that the particle part
and the antiparticle part of the Hamiltonian separate.
This enables them to derive relatively simple analytical expressions
for the quasiparticle energies. Although in general these solutions
are not exact (corresponding to the zeros of the polynomials in our 
\eq{trlog}),
the numerical results obtained in this way look very similar to ours,
thereby justifying the above approximation.}\label{foot2} 

The Matsubara sums over \eqs{trlog2cs} and (\ref{trlog3cfl}) can again 
be turned out with the help of \eq{Matsu}.
In general, i.e.,  for \eq{trlog} with arbitray values of the 
condensates, this cannot be done so easily. 
If one combines the second with the third logarithm on the r.h.s., the
argument becomes a polynomial of fourth order in $\omega_n^2$. 
The same is true for the argument of the fourth logarithm. 
The corresponding dispersion laws are related to the zeros of these
polynomials. Although, in principle, the zeros of a polynomial of fourth 
order can be determined analytically, the results are very nasty expressions
which are difficult to handle. Therfore, in practice one has to determine
the dispersion laws numerically. After that, one can again employ \eq{Matsu} 
to calculate the Matsubara sum.  
Alternatively, one can turn out the Matsubara sum numerically without
previous determination of the dispersion laws. This is the method we have
used.

So far, the thermodynamic potential depends on our choice of the condensates
$\phi_u$, $\phi_s$, $\ct$, and $\cf$. On the other hand, in a 
thermodynamically consistent treatment the condensates should follow 
from the thermodynamic potential by taking the appropriate derivatives.
The self-consistent solutions are given by the stationary points of the
potential,
\beq
    \frac{\delta \Omega}{\delta \phi_u} \= 
    \frac{\delta \Omega}{\delta \phi_s} \=
    \frac{\delta \Omega}{\delta \ct} \=
    \frac{\delta \Omega}{\delta \ct^*} \=
    \frac{\delta \Omega}{\delta \cf} \=
    \frac{\delta \Omega}{\delta \cf^*} \= 0 \,.
\end{equation}
In this way one finds that the quark-antiquark condensates are given by
\begin{alignat}{2}
    \phi_u \=& &-\frac{1}{2}\,&T \sum_n \int \frac{d^3p}{(2\pi)^3} \;
    \frac{\partial}{\partial M_u}\; 
    \frac{1}{2}\,{\rm Tr}\;\ln\, \Big(\frac{1}{T}\,S^{-1}(i\omega_n, \vec p)
    \Big)\,,
\nonumber \\
    \phi_s \=& &-&T \sum_n \int \frac{d^3p}{(2\pi)^3} \;
    \frac{\partial}{\partial M_s}\; 
    \frac{1}{2}\,{\rm Tr}\;\ln\, \Big(\frac{1}{T}\,S^{-1}(i\omega_n, \vec p)
    \Big)\,.
\label{gapphi}
\end{alignat}
The explicit evaluation of the derivatives is trivial, but 
leads to rather lengthy and not very illuminating expressions, which we 
do not want to present. 
Note, however, that $M_u$ and $M_s$, which determine the right hand sides,
depend again on $\phi_u$ and $\phi_s$ (see \eq{Mus}) and therefore the
equations have to be solved self-consistently.
 
Similarly one finds that the diquark condensates are governed by the
gap equations
\begin{alignat}{2}
    \Delta_2 \=& &4H\Delta_2\,&T \sum_n \int \frac{d^3p}{(2\pi)^3} \;
    \frac{\partial}{\partial\dt^2}\; 
    \frac{1}{2}\,{\rm Tr}\;\ln\, \Big(\frac{1}{T}\,S^{-1}(i\omega_n, \vec p)
    \Big)\,,
\nonumber \\
    \Delta_5 \=& &2H\Delta_5\,&T \sum_n \int \frac{d^3p}{(2\pi)^3} \;
    \frac{\partial}{\partial \df^2}\; 
    \frac{1}{2}\,{\rm Tr}\;\ln\, \Big(\frac{1}{T}\,S^{-1}(i\omega_n, \vec p)
    \Big)\,.
\label{gapdelta}
\end{alignat}
Obviously, both equations have trivial solutions $\Delta_A=0$ which are
independent of each other. For the nontrivial solutions \eqs{gapphi} and 
(\ref{gapdelta}) are coupled and have to be solved simultaneously.
If there is more than one solution, the stable solution is the 
one which corresponds to the lowest value of $\Omega$.

\section{Numerical results}
\label{results}   

For the numerical evaluation of the equations derived in the previous 
section we first have to specify the interaction. 
As a typical example we consider a four-fermion interaction with the quantum numbers of a
single-gluon exchange,
\beq
    {\cal L}_{int} \= -g\,\sum_{a=1}^8 (\pb \gamma^\mu \lambda_a\psi)^2 \,.
\label{Lint}
\end{equation}         
This interaction was also the starting point of the model calculations 
in Refs.~\cite{ARW99} and \cite{ABR99}. Performing Fierz transformations we 
obtain for the effective coupling constants which enter \eqs{Lqbarq} and 
(\ref{Lqq})
\beq
     G \= \frac{8}{9}\,g \;, \qquad  
     H \= \frac{2}{3}\,g \;.
\label{Fierz}
\end{equation}
In addition, ${\cal L}_{int}$ gives rise to effective
couplings in other channels, which in principle have to be taken into
account to be fully self-consistent in the sense that also all
exchange \mbox{(Fock-) terms} are included. In fact, already for two  
flavors, one finds that a self-consistent treatment requires the simultaneous 
consideration of possible expectation values in six different channels, 
like color dependent quark condensates or densities \cite{BHO01}.
For three flavors, there are even more. We already mentioned the induced
color-sextet diquark condensates in the CFL-phase \cite{ARW99,Sch00,Sho99}. 
However, most of these induced condensates -- as far as they have been
studied so far -- have turned out to be small. Therefore, since the
structure of the equations is rather involved even without these
additional channels, we neglect them in this paper. Also, as stated above,
\eq{Lint} should only be viewed as a ``typical''
interaction, which we have chosen for a qualitative illustration of the
more general equations derived in the previous section.
Certainly, there is no reason to exclude other terms
(see also the comments in Ref.~\cite{ABR99} about this point).

Taking the model as defined above, there are four parameters: the
coupling constant $g$, the bare quark masses $m_u$ and $m_s$, and a
cut-off parameter, which is needed to regularize the divergent
momentum integrals in \eqs{Omega}, (\ref{gapphi}) and
(\ref{gapdelta}). For simplicity, we take a sharp 3-momentum cut-off
$\Lambda$. We expect that the results will be qualitatively the same,
if a smooth form factor is used.  As customary (see e.g.\@ the
comments on this point in Ref.~\cite{BeRa99}), the parameters are
chosen such that ``reasonable'' vacuum properties are obtained. We
take $\Lambda = 600$~MeV, $g\Lambda^2 = 2.6$, $m_u = 5$~MeV and $m_s =
120$~MeV. With these parameters we find the vacuum constituent quark
masses $M_u = 362.5$~MeV and $M_s =557.4$~MeV, which corresponds to
the condensates $\phi_u = (-240.5 {\rm MeV})^3$ and $\phi_s = (-257.3
{\rm MeV})^3$. These values are similar to those obtained in
Ref.~\cite{Rehberg}, where the parameters have been fixed by fitting 
vacuum masses and decay constants of pseudoscalar mesons. 

We begin with the discussion of the results at zero temperature.  
\begin{figure}
\begin{center}
\epsfig{file=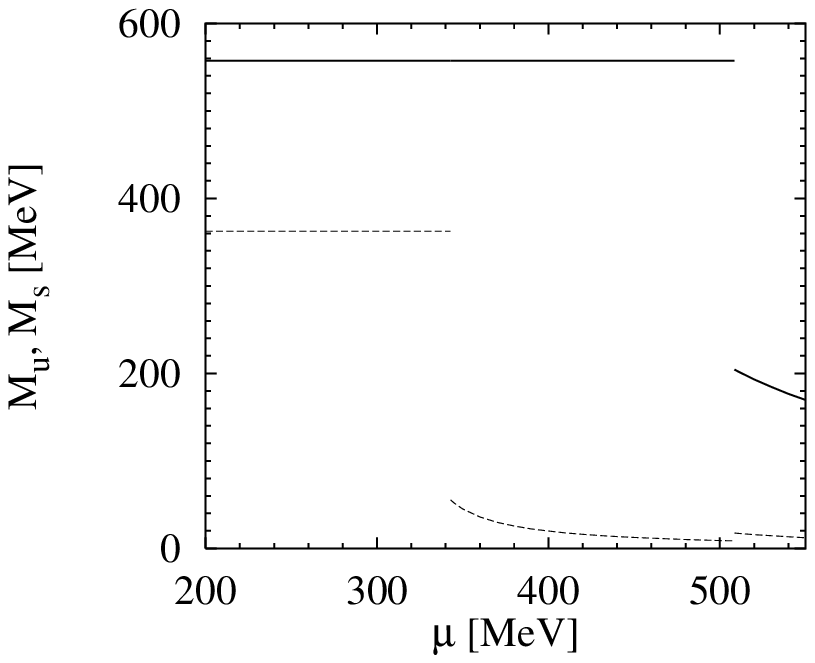,width=7.5cm}
\hfill
\epsfig{file=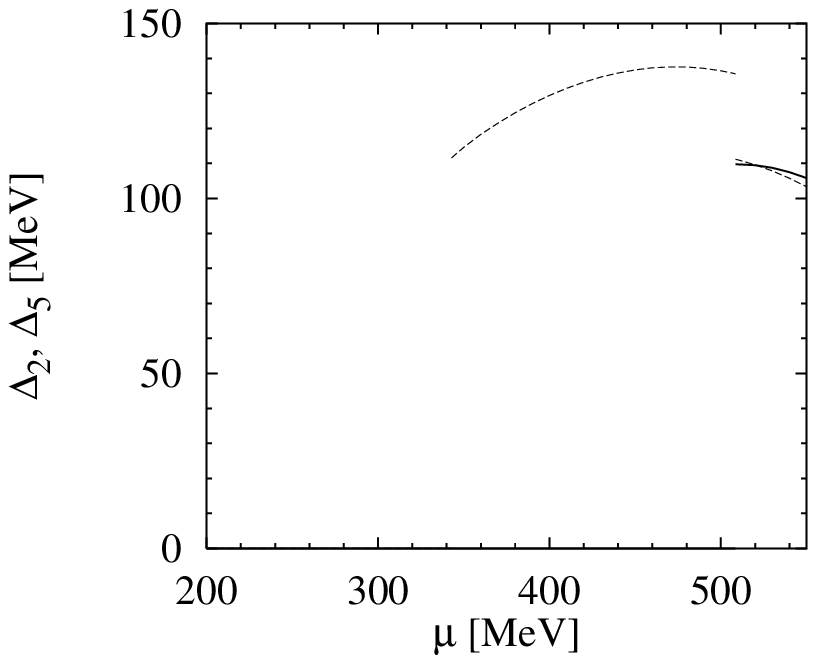,width=7.5cm}
\end{center}
\vspace{-0.5cm}
\caption{\small Gap parameters at $T=0$ as functions of the quark chemical 
          potential $\mu$.
          Left: Constituent masses of up and down quarks (dashed), 
          and of strange quarks (solid).
          Right: Diquark gaps $\Delta_2$ (dashed) and $\Delta_5$ (solid).}
\label{figt=0}
\end{figure}
The behavior of the four gap parameters as functions of the quark chemical
potential $\mu$ is displayed in Fig.~\ref{figt=0}. In the left panel we
show the constituent quark masses $M_u$ and $M_s$, in the right panel the
diquark gaps $\Delta_2$ and $\Delta_5$.
One can clearly distinguish three phases. 
At low chemical potentials $\mu < \mu_1 =342.8$~MeV, the diquark gaps 
vanish and the constituent quark masses stay at their vacuum values.
For the condensates this means $\ct = \cf = \cs = 0$, while
$\phi_u$ and $\phi_s$ are large. Hence, in a very schematic sense, this
phase can be identified with the ``hadronic phase'', although there are
no hadrons in our model. In fact, in Fig.~\ref{figt=0} the entire region
below $\mu=\mu_1$ corresponds to the vacuum solution with zero density. 

At $\mu = \mu_1$ a first-order phase transition takes place and the 
system becomes a two-flavor color superconductor:
The diquark condensate $\ct$ has now a non-vanishing expectation value,
related to a non-vanishing diquark gap $\Delta_2$, whereas $\Delta_5$ 
remains zero.
Just above the phase transition we find $\Delta_2 = 111.5$~MeV. At the
same time the mass of the up quark drops from the vacuum value to 
$M_u = 55.3$~MeV. With increasing $\mu$, $M_u$ decreases further, while
$\Delta_2$ increases until it reaches a maximum at $\mu \simeq 475$~MeV. 
Just below the next phase transition at $\mu = \mu_2 = 508.5$~MeV we find 
$\Delta_2 = 135.6$~MeV and $M_u = 8.6$~MeV. 

In the 2SC phase the baryon number density is no longer
zero and increases from $\rho_B = 0.42\; {\rm fm}^{-3}$ (2.5 times nuclear 
matter density) above $\mu = \mu_1$ to $\rho_B = 1.15\; {\rm fm}^{-3}$ 
(6.7 times nuclear matter density) below $\mu = \mu_2$. 
The density of strange quarks remains zero up to $\mu = \mu_2$, but this
could be an artifact of the interaction which contains no flavor mixing
terms. As a consequence the strange quark mass $M_s$ remains at its vacuum
value and hence strange quark states cannot be polulated in this regime. 
A flavor mixing interaction would lead to a reduced value of $M_s$ in the 
2SC phase and thus to a lower threshold for a finite density of strange 
quarks.

At $\mu = \mu_2$ the system undergoes a second phase transition, this time
from the 2SC phase into the CFL phase, which is characterized by a 
non-vanishing diquark gap $\Delta_5$ (together with a non-vanishing
$\Delta_2$). The phase transition is again of first order: At the
transition point $\Delta_5$ jumps from zero to 109.7~MeV, while $M_s$
drops from 557.4~MeV to 204.3~MeV. The non-strange quantities are also
discontinous and change in the opposite direction: $\Delta_2$ drops
from 135.6~MeV to 111.2~MeV, and $M_u$ jumps from 8.6~MeV to
17.5~MeV. The density jumps from  $\rho_B = 1.15\; {\rm fm}^{-3}$ to
$1.63\; {\rm fm}^{-3}$,
i.e., from 6.7 to 9.6 times nuclear matter density. 

If we started from an exact SU(3) flavor symmetry, i.e., equal masses
for all quark flavors, we would expect $\Delta_2 = \Delta_5$ in the CFL
phase. However, as we are far away from an exact symmetry, 
it is remarkable that the diquark gaps $\Delta_2$ and $\Delta_5$ are so
similar at the transition point. 
The reason for the phase transition is not that $u$ and $s$
quarks can no longer condense below that value. 
We still find a 
CFL solution down to $\mu = 445.0$~MeV. However, for $\mu < \mu_2$, 
there is a more favored solution with a much higher strange quark mass which 
cannot support a condensation of $u$ and $s$ quarks. In this regime
the CFL solution corresponds to a metastable state. 

In Ref.~\cite{ABR99} it has been argued that the
color-flavor-unlocking transition at zero temperature has to be first
order because pairing between light and strange quarks can only occur
if the gap is of the same order as the mismatch between the Fermi
surfaces. Hence the value of the gap must be discontineous at the 
phase transition. The existence of pairing between light and strange 
quarks does of course not automatically mean that the CFL solution
corresponds to an absolutely stable solution. This has already been 
emphasized in Refs.~\cite{RaWi0012,ARRW01}. The quantitative criteria 
derived in these references for the existence of a CFL solution as a
global minimum of the thermodynamic potential cannot be applied to our 
case because the possibility of a density dependent strange  quark mass 
has not been taken into account in the derivation.
However, at least qualitatively, our results support the arguments put 
forward in these references: There is indeed a
non-vanishing lowest possible value of $\Delta_5$ for metastable CFL 
solutions and the minimal value of $\Delta_5$ in an absolutely stable 
CFL phase is even larger. 

At large $\mu$ we become more and more sensitive to the cut-off and we
approach the limits of the model. This is also the reason why at large 
chemical potentials both diquark condensates decrease with $\mu$. 
Although it is sometimes argued that the cut-off simulates asymptotic
freedom, i.e., the weakening of the strong coupling constant at large
(Fermi-) momenta in some crude way, the results should 
certainly not been taken too serious in this regime. In particular the
fact that we find $\Delta_5 > \Delta_2$ above $\mu \simeq 520$~MeV should
be seen as a cut-off artifact. 

\begin{figure}
\begin{center}
\epsfig{file=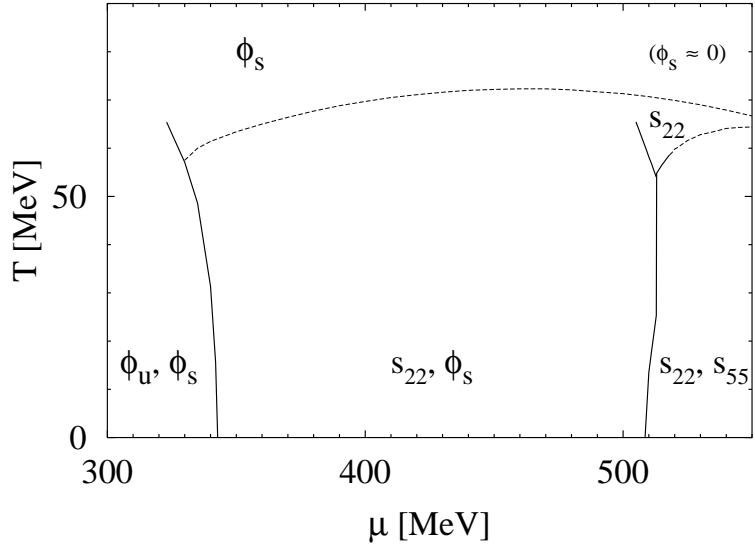,width=11.0cm}
\end{center}
\vspace{-0.5cm}
\caption{\small Phase diagram in the $\mu-T$ plane. The solid lines 
         indicate phase boundaries with a first-order phase transition,
         the dashed lines correspond to second-order phase transitions.
         The different phases can
be distuingished by different values for the various
condensates. Within the figure we have indicated only those condensates
which are significantly different from zero. (Note that because of isospin
symmetry, $\phi_d = \phi_u$ and $s_{77} = s_{55}$.) 
}
\label{figphase}
\end{figure}

We now extend our analysis to non-vanishing temperatures. The
resulting phase diagram in the $\mu-T$ plane is shown in
Fig.~\ref{figphase}. Let us first concentrate on the quark-antiquark
condensates $\phi_u$ and $\phi_s$. We could then distinguish between three 
different regimes: At low temperatures and low chemical potentials both
condensates, $\phi_u$ and $\phi_s$ are relatively large and we can
schematically identify this phase with the hadronic phase, as before.
When we increase the temperature or the chemical potential, eventually
a second phase is reached, where $\phi_u$ is almost zero. $\phi_s$ is still 
large in this phase and it becomes small only in a third phase at even 
higher temperatures or chemical potentials. At low temperatures, e.g., 
along the T=0-axis, these three phases are well separated by first-order 
phase transitions, whereas at high temperatures we only find smooth 
cross-overs. Consequently, both first-order phase transition lines end in a 
second-order endpoint, which in our model is located at $T \simeq 66$~MeV and
$\mu \simeq 322$~MeV for the phase transition related to $\phi_u$ and at
$T \simeq 66$~MeV and $\mu \simeq 505$~MeV for the phase transition related 
to $\phi_s$.
For the former we should keep in mind, however, that our model does neither 
contain hadronic nor gluonic degrees of freedom, which are certainly important
in this area of the phase diagram.

We now turn to the diquark condensates.
As discussed for the zero temperature case, at low temperatures, the three
phases discussed in the previous paragraph coincide with the ``hadronic
phase'', the 2SC phase and the CFL phase, i.e., we have
$\ct = \cf = 0$ at low chemical potentials, $\cf = 0$,
but $\ct \neq 0$ at intermediate chemical potentials, and 
$\ct \neq 0$ and $\cf \neq 0$ at high chemical potentials.
At high temperatures all diquark condensates vanish. This leads to two
additional phase transition lines.   
The first one corresponds to the transition from the 2SC phase to the
``quark-gluon'' plasma, i.e., to the phase with vanishing diquark condensates
and (almost) vanishing $\phi_u$. In agreement with various BCS-type
calculations, this phase transition is of second order. In addition
we obtain almost perfect agreement with a universal
relation which can be derived (see e.g.\@ \cite{FW71,PiRi99})
between the value of the gap parameter (in our case this corresponds
to $\Delta_2$) at zero temperature and the critical temperature $T_c$
for that transition:
\beq
T_c \approx 0.57 \Delta_2 (T = 0)~.
\label{Tcapp}
\end{equation}
For instance, at $\mu = 343$~MeV we find $\Delta_2(T = 0 ) = 111.5$~MeV,
and $T_c = 62.0$~MeV, whereas from the above relation we would expect
$T_c = 63.5 $~MeV. Since \eq{Tcapp} is an approximate relation, 
based on the assumption that pairing takes place only in the vicinity
of the 
Fermi surface, the agreement is quite good.

\begin{figure}
\begin{center}
\epsfig{file=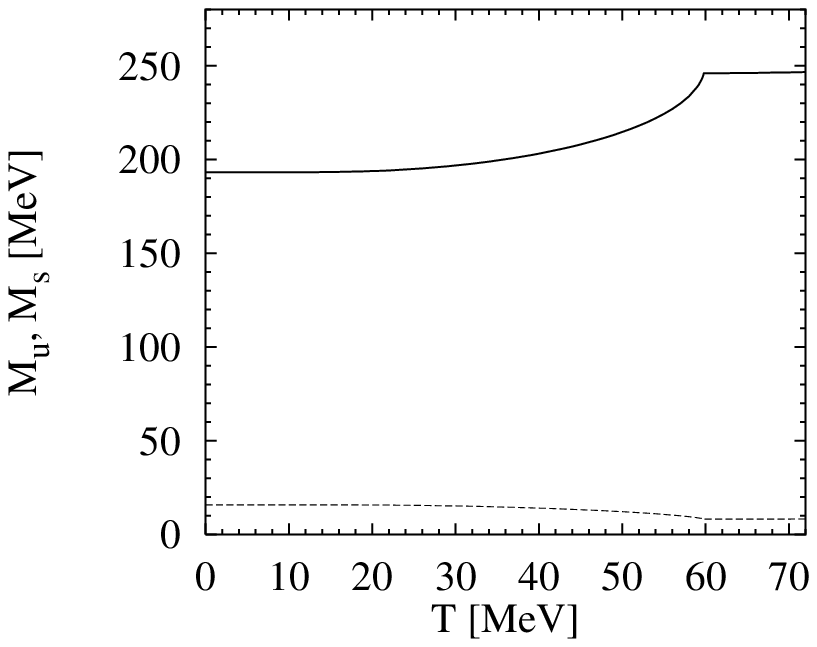,width=7.5cm}
\hfill
\epsfig{file=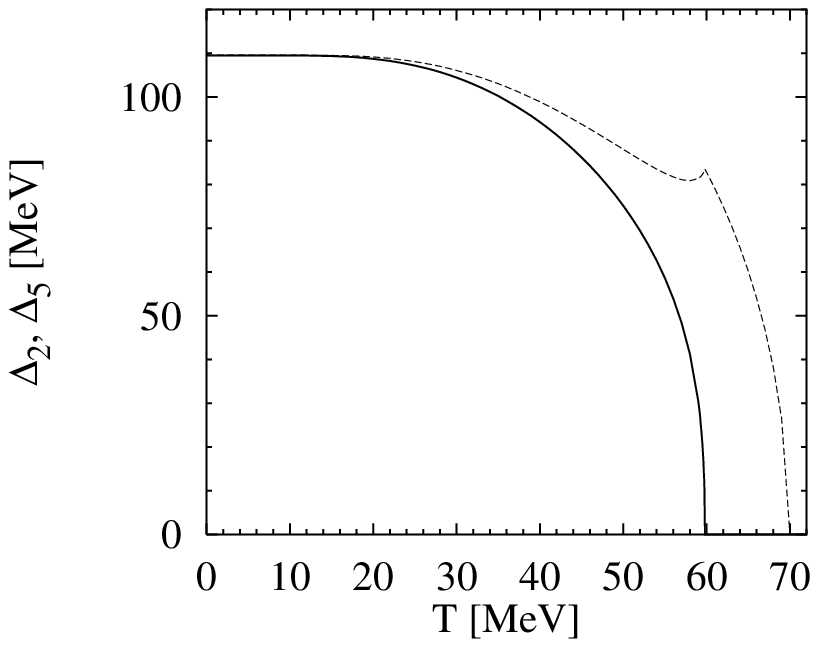,width=7.5cm}
\end{center}
\vspace{-0.5cm}
\caption{\small Gap parameters at $\mu = 520$~MeV as functions of the 
          temperature.
          Left: Constituent masses of up and down quarks (dashed), 
          and of strange quarks (solid).
          Right: Diquark gaps $\Delta_2$ (dashed) and $\Delta_5$ (solid).}
\label{figmu=520}
\end{figure}
 
Starting from the CFL phase and increasing the temperature, one first 
observes a ``melting'' of the diquark gap $\Delta_5$, before at
somewhat higher temperatures $\Delta_2$ also vanishes. The intermediate
2SC phase ``above'' the CFL-phase is distinguished from the regime
``left'' to the CFL-phase by a much smaller value of $\phi_s$ and hence
of the strange quark mass.  
Based on the assumption that $\Delta_5$ is always smaller than $\Delta_2$
(which is not really true at $T=0$, as we have seen), the earlier 
disappearance of $\Delta_5$ and hence the existence of a 2SC phase ``above''
the CFL phase was already anticipated in Ref.~\cite{RaWi00}. 
Applying similar arguments as at zero temperature 
(cf.~Refs.~\cite{ABR99,RaWi0012,ARRW01}) it was also predicted in that 
reference, that the corresponding color-flavor unlocking transition should 
again be of first order. This was corrobated by a second argument, claiming 
that the phase transition
corresponds to a finite temperature chiral restoration phase transition 
in a three-flavor theory. In this case the universality arguments of 
Ref.~\cite{PiWi84} should apply, stating that the phase transition 
should be of first order. 

Indeed, following the phase transition line from the left, 
we find that the transition continues to be first order. 
However, above a tricritical point at $\mu \simeq 518$ MeV and 
$T \simeq 59$ MeV the phase transition becomes second order.
This is illustrated in Fig.~\ref{figmu=520}, where the
constituent masses (left panel) and the diquark gaps (right panel)
are displayed as functions of the temperature for fixed $\mu = 520$~MeV.  
We find a second-order phase transition at $T = 59.8$~MeV.
Of course we cannot exclude that the second-order phase transition is 
just an artifact of our simplified model or the present approximation. 
On the other hand, the arguments of Refs.~\cite{ABR99,RaWi0012,ARRW01} in 
favor of a first-order phase transition are much less stringent at finite
temperature, where even for vanishing condensates the Fermi surfaces are 
smeared due to thermal effects. The applicability of the universality
argument is also questionable in the present situation because
the 2SC phase is not a three-flavor chirally
restored phase, but only $SU(2)\times SU(2)$ symmetric. Note that even 
$SU(3)_V$ is broken explicitly by unequal current quark masses and 
spontaneously, e.g., by the flavor-antitriplet diquark
condensates. Therefore a rigorous prediction of the true order of the
phase transition is rather difficult.  

Finally, we should repeat that our results at $\mu \gsim$~500~MeV 
are likely to be sensitive to the cut-off. Therfore it remains to be
studied in a more systematic way, which of our findings persist 
if the model parameters (including the cut-off) are varied over a wider 
range. Work in this direction is in progress.

\section{Conclusions}
\label{conclusions}
Recent investigations revealed a rich phase structure of strongly
interacting matter. Besides the various known phases of hadronic
matter and the quark-gluon plasma phase at high temperatures several
``superconducting'' phases are likely to exist in cold dense
matter. Calculations within effective models showed that in this
context ``dense'' could mean only a few times nuclear matter density
such that those superconducting phases could be relevant for the
interior of neutron stars or even for relativistic heavy ion collisions
\cite{RaWi00}. In this paper we focused on the study of this ``cold
dense'' region within the framework of an NJL-type effective quark
model. This model allows for condensation in various channels, in
particular chiral symmetry breaking due to quark-antiquark
condensation and diquark condensation can be described. Our main
intention was to examine the effect of selfconsistently (via
$\bar{q}q$-condensation) determined realistic quark masses. In
particular the mass of the strange quark plays an important role. For
the two extreme cases, taking the strange quark mass equal to the
light quark masses or considering an infinite strange quark mass, two
different condensation patterns are found, respectively: The so-called
color-flavor-locked phase \cite{ARW99}, where all quarks contribute to 
diquark condensation and the two-flavor color superconductor
\cite{ARW98,RSSV98}, where only light quarks participate in the condensation. 
We could show that both limiting cases are correctly described within our
model. However, as in the regions of interest the strange quark mass is 
neither negligible nor infinite the realistic situation must be somewhere
in between. Indeed, for quark matter at zero temperature, we find that for 
moderate values of the chemical potential the two-flavor superconductor is 
energetically favored, corresponding to densities of about 2.5 $\rho_0$ up 
to about 6.7 $\rho_0$.  Only for even higher densities the CFL phase is found
to be the energetically most favored one. The main reason is the large
constituent mass of the strange quarks compared with that of up and
down quarks which has been determined selfconsistently within our
investigations. In particular, the strange quark mass in the 2SC phase is 
much larger than than in the CFL phase  and the phase transition is mostly
triggered by a discontinous change of the quark mass. 

The situation remains qualitatively unchanged up to temperatures of about 
50 MeV. At higher temperatures the diquark condensates melt. For the diquark
condensates which involve strange quarks this happens at somewhat lower
temperatures than for the non-strange condensate. Hence, with increasing
temperature, the system passes from the CFL phase through an intermediate
2SC phase to the quark-gluon plasma. 
In contrast to the situation at zero
temperature, we found that the strange quark mass in this
``high-temperature''
2SC phase is comparable to the strange quark mass in the CFL phase.
Furthermore we found a tricritical point on the color-flavor unlocking
line
above which the phase transition is second order.

Of course, our model is still rather schematic and 
can only give some qualitative hints on the interdependencies between the 
effective quark masses and the various superconducting phases. 
Unfortunately, as our knowledge about the effective quark interaction 
to be used at moderate chemical potentials is limited,
more quantitative predictions are difficult. In this situation a systematic
examination of the sensitivity of the results on the choice of the 
interaction might be helpful to sort out some common features.
Perhaps the most severe simplification in our model was that we did not 
include a 't Hooft-type interaction, which mixes different
flavors and thus affects the strange quark mass even in regimes, where no 
strange quark states are populated. It is therefore
certainly worthwhile to investigate how such a term influences the phase 
structure. In a more systematic analysis one should also study the effect 
of other condensates, which have been neglected so far. 

For simplicity, we restricted our studies to a common baryon chemical 
potential for all flavors. However, for many applications, 
like quark cores of neutron stars or the possible existence of absolutely
stable strange quark matter, it would be interesting
to consider $\beta$-equillibrated quark matter, where the chemical
potential for up quarks is in general different from that of down and
strange quarks. 
This could alter the position of the various Fermi surfaces and hence
the entire phase structure. In an earlier work without diquark condensation
we have shown that a selfconsistent treatment of the strange quark constituent
mass is a very important effect in this context \cite{BuOe99}. 
On the other hand, in more recent studies which include diquark condensation
(e.g., about the existence of the color-flavor locked phase in neutron 
stars \cite{Reddy} or strangelets \cite{Madsen01}) the interdependencies
between strange quark mass and phase structure have so far been neglected.
In view of our present results, work in this direction seems to be worth 
being further pursued.

\section*{Acknowledgments}
We thank Krishna Rajagopal for useful comments and discussions.
This work was initialized during the collaboration meeting on
color superconductivity in Trento. We thank ECT$^*$ for financial support
during this meeting and the organizer, Georges Ripka, and all
participants, in particular Ji\v ri Ho\v sek and Maria-Paola Lombardo,
for stimulating discussions. We also thank Ji\v ri Ho\v sek for his
hospitality at the Nuclear Physics Institute in \v Re\v z. 
One of us (M.O.) acknowledges support from the Alexander von
Humboldt-foundation as a Feodor-Lynen fellow. This work was in part
supported by the BMBF.

\end{document}